\documentstyle[aps,multicol,epsf]{revtex}
\begin{document}
\preprint{}
\draft
\title{General Localization Lengths for Two Interacting Particles in a
Disordered Chain}
\author{Pil Hun Song and Felix von Oppen}
\address{Max-Planck-Institut f\"{u}r Kernphysik, Postfach 103980,
69029 Heidelberg, Germany} 
\date{\today}
\maketitle
\begin{abstract}
The propagation of an interacting particle pair in a disordered chain
is characterized by a set of localization lengths which we define. 
The localization lengths are computed by a new decimation
algorithm and provide a more comprehensive picture of the two-particle 
propagation.  We find that the interaction delocalizes predominantly
the center-of-mass motion of the pair and use our approach to
propose a consistent interpretation of the discrepancies between
previous numerical results.
\end{abstract}


\begin{multicols}{2}
\narrowtext

The problem of interacting electrons in a disordered potential is one
of the important unsolved problems in condensed matter physics.  This
has been emphasized again by the recent observation\cite{krav} of a
metal-insulator transition in two dimensional (2d) systems which was
theoretically unanticipated.  Some time ago, Shepelyansky\cite{she} 
proposed that it would be worthwhile to consider the simple case of two
interacting particles in a random potential. He predicted that
unexpectedly, such a particle pair could propagate coherently over
distances $\xi_2$ much larger than the single-particle localization
length $\xi_1$ as long as the two particles are within $\xi_1$ from
each other.

Specifically, Shepelyansky obtained for the two-particle localization
length \begin{equation}
    \xi_2 \sim (U/W^2)^2, 
\end{equation} 
where $U$ denotes the interaction strength and $W$ the disorder
strength. Since $\xi_1\sim 1/W^2$, Eq.~(1) implies an enhancement of
the localization length for weak disorder. Shepelyansky's original
argument involved several uncontrolled assumptions for the
single-particle eigenstates. This led to a number of (mostly
numerical) attempts\cite{Imr,fra1,opp,song,rom} to study the problem of
two interacting particles more rigorously. Imry \cite{Imr} rederived
Shepelyansky's result, Eq.~(1), by an extension of the Thouless block scaling
picture. Frahm {\it et al.}  \cite{fra1} computed $\xi_2 \sim
W^{-3.3}$ using the transfer matrix method (TMM). Von Oppen {\it et
al.}\cite{opp} introduced a Green function approach, allowing one to
project the problem on the subspace of doubly occupied sites, and
concluded $\xi_2 \sim U/W^4$.  Subsequently, Song and Kim\cite{song}
treated the idea of von Oppen {\it et al.} rigorously using the
recursive Green function method and found $\xi_2 \sim W^{-2.9}$.

Recently, R\"omer and Schreiber\cite{rom} concluded from the TMM that
the enhancement effect does {\it not} exist. In view of this claim and of the
quantitatively different expressions for $\xi_2$ quoted above, it
appears that there are few secured results in this field. Our purpose
in this paper is to present a more comprehensive picture of the
two-particle propagation by defining and computing a set of
localization lengths. We unambiguously show that the effect exists and
propose a resolution of the controversy in the previous works\cite{rom,fra2}.

It is currently not clear whether these ideas have any relevance to
the degenerate finite-density Fermi gas. It appears to be the most
promising direction to consider the localization properties of
quasiparticle pairs. There have been a number of
studies\cite{Imr,opp2,she2} whether quasiparticle excitations
delocalize relative to single-particle ones.  While a numerical study
for a one-dimensional (1d) system showed delocalization only for
unrealistically high excitation energy of the pair (of order of the
bandwidth)\cite{opp2}, both arguments\cite{Imr} and numerical 
studies\cite{she2} in higher 
dimensions suggest the possibility of a new pair mobility edge close to the
ground state.

The two-particle problem in one-dimension is described by the
Hamiltonian
\begin{equation}
\label{eq-hamil}
{\cal H} = H_1 \otimes {\bf 1} + {\bf 1} \otimes H_1 +
U \sum_m |m\rangle|m\rangle \langle m| \langle m|,
\end{equation}
where $m$ labels the $N$ sites of the 1d lattice and $H_1$ is the
usual single-particle Anderson Hamiltonian
\begin{equation}
H_1 = \sum_m [\ \epsilon_m|m\rangle \langle m| + t\ (|m\rangle \langle
m+1|+|m+1\rangle \langle m|)\ ].
\end{equation}
$\epsilon_m$ is a random site energy, drawn from a box distribution
with $-W/2\!\leq\!\epsilon_m\!\leq\! W/2$, and $U$ the on-site
interaction.  The hopping matrix element $t$ is set to unity throughout 
this work.  A convenient quantity to study the localization properties
of the pair is the two-particle Green function $G=(E-{\cal H})^{-1}$.
The two-particle localization length $\xi_2$ on which previous studies 
have focused is defined in terms of $G$ as \cite{opp}
\begin{equation}
\label{xi2-green}
  \xi_{2}^{-1} = -\lim_{|n-m| \rightarrow \infty}\frac{1}{|n-m|}
  \ll\ln |\langle m,m|G|n,n\rangle|\gg,
\end{equation}
where the double bracket denotes the disorder average.  

In this paper, we discuss general localization lengths which
provide a much more comprehensive picture of the localization properties of
the particle pair.  First, we consider a general center-of-mass (CM)
motion by defining
\begin{equation}
  \xi_{2,a}^{-1} = -\lim_{|n-m| \rightarrow \infty}\frac{1}{|n-m|}
  \ll\ln |\langle m,m-a|G|n,n-a\rangle|\gg.  \end{equation} We find
that, surprisingly, $\xi_{2,a}$ is essentially independent of the
particle distance $a$, even if $a$ exceeds the single-particle
localization length $\xi_1$. We also study the behavior of $G$ for 
relative motion at fixed CM, as characterized by

\begin{eqnarray}
  \xi_r^{-1} &=& -\lim_{n \rightarrow \infty}\frac{1}{n} \ll\ln |\langle
   m+n,m-n|G|m,m\rangle|\gg.  
\end{eqnarray} 
Finally, we consider the propagation of one of the particles with the
other held fixed, as described by
\begin{eqnarray}
  \xi_f^{-1} &=& -\lim_{n \rightarrow \infty}\frac{1}{n}
  \ll\ln |\langle m,m+n|G|m,m\rangle|\gg.  
\end{eqnarray} 
As opposed to $\xi_2$ and $\xi_{2,a}$, we find that the latter two
lengths are only very weakly affected by the interaction $U$.
Nevertheless, it will turn out that these lengths are indispensable for
obtaining a more comprehensive picture of the two-particle propagation and
for understanding the discrepancies between previous numerical results.

While $\xi_2$ could be computed by projecting the problem on the 
subspace of doubly occupied sites, this is no longer possible
for the generalized localization lengths defined above. For this
reason, we introduce a new decimation algorithm, which allows us to
compute these localization lengths efficiently. As opposed to the
projection method for $\xi_2$ used in ref.\ \onlinecite{opp}, this
algorithm is numerically exact.  We briefly describe the procedure for
computing $\xi_2$.  Adaption to the other lengths defined above is
straightforward. Since the interaction acts only on symmetric states,
we specify to (spinless) bosons.  Using a symmetrized basis 
\begin{equation}
  \label{eq-basis} |mn\rangle = \left\{ \begin{array}{ll}
  |m\rangle|m\rangle & \mbox{\ \ \ \ if $m = n$}, \\
  (1/\sqrt{2})(|m\rangle|n\rangle+|n\rangle|m\rangle) & \mbox{\ \ \ \ if
  $m \neq n$}.  \end{array} \right.  
\end{equation} 
and interpreting
$(m,n)$ as sites of a 2d square lattice, the Hamiltonian of
Eq.~(\ref{eq-hamil}) can be interpreted as describing a single
particle on the 2d lattice shown by the thin solid lines in Fig.~1.
The off-diagonal elements of ${\cal H}$ are nonzero only for
nearest-neighbor bonds and equal to $\sqrt{2}$ (1) if one (none) of
the nearest-neighbor sites is a doubly occupied state. Our goal is to compute
the Green function $G(E)$ which is the inverse of a sparse matrix,
$D=E-{\cal H}$ of linear size $\sim N^2$.  Clearly, a direct
manipulation of the whole matrix is inefficient both in terms of time
and storage, and becomes forbidding for $N>100$. To circumvent this
problem, we recursively decimate the irrelevant matrix elements of the
Green function. We start by decomposing Hilbert space into subspaces
$i$, each of which is spanned by the states along one of the dashed
lines in Fig.~1(a) and which are labeled by their dimensions $1\leq
i\leq N$. We denote the projection of $D$ onto these subspaces as
$D_i$. Clearly, $D$ couples only neighboring subspaces $(i)$ and
$(i+1)$, and we call the corresponding ($i \times (i+1)$ dimensional)
coupling block in the Hamiltonian $V_i$. Finally, we define vectors
${\bf x}_i^{(n)}$ with elements 
\begin{equation} ({\bf x}_i^{(n)})_j = \langle N-i+j,j|G(E)|nn\rangle, 
\end{equation} 
given by matrix elements
of the Green function $G$ between a doubly occupied site $|nn\rangle$
and the states in subspace $i$. Since only neighboring subspaces are
coupled, one readily derives from $DG=1$ the set of coupled linear
equations 
\begin{eqnarray}
 D_1 {\bf x}^{(n)}_1 + V_1 {\bf x}^{(n)}_2 &=& 0, \nonumber \\
 V_{i-1}^T {\bf x}^{(n)}_{i-1} + D_i {\bf x}^{(n)}_i + 
 V_i {\bf x}^{(n)}_{i+1} &=& 0, \ \ \ \ 2\leq i\leq N-1, \nonumber \\
 V_{N-1}^T {\bf x}^{(n)}_{N-1} + D_N {\bf x}^{(n)}_N &=& {\bf e}_n,
\end{eqnarray}
where ${\bf e}_n$ is the $N$ dimensional unit vector with
$({\bf e}_n)_m=\delta_{n,m}$.  Solving these equations, we obtain
\begin{equation}
  {\bf x}^{(n)}_N = {\cal G}_N {\bf e}_n,
\end{equation}
where the ${\cal G}_i$ can be computed recursively from
\begin{equation}
{\cal G}_i= (D_i - V_{i-1}^T {\cal G}_{i-1} V_{i-1})^{-1} \ \ \ 
\mbox{with ${\cal G}_1 = D_1^{-1}$}. 
\end{equation}
Finally noting that $({\bf x}^{(n)}_N)_l=\langle ll|G(E)|nn\rangle$,
we can now compute the localization length $\xi_2$ from Eq.\ 
(\ref{xi2-green}). This reduces the
calculation to manipulations of matrices of sizes from $1 \times 1$ to $N
\times N$. It is worthwhile to point out that at the final stage of
the iteration, the calculation is formally reduced to an effective 1d
model for a single particle. It is straightforward to generalize the
algorithm to compute the other localization lengths defined in this
paper.  E.g., $\xi_r$ is calculated by decomposing Hilbert space
according to the dotted lines of Fig.~1(a).
 
For $\xi_2$, we set $n=1$ in the above algorithm and obtain $t_{l,1} =
\ \ll\ln|\langle ll|G|11\rangle |\gg$ with $1\leq l\leq N$ for each
parameter set $(W,U)$.  We find that $t_{l,1}$ depends linearly on $l$, 
implying an exponential decay of the Green function.  To
eliminate finite size effects near $l=1$ and $l=N$, we fit $t_{l,1}$
in the range $N/5 \leq l \leq 4N/5$ to 
\begin{equation}
t_{l,1} = -\frac{l}{\xi_2} + c
\end{equation}
with $c$ a constant. We find that for chains $N\geq200$, our results
for $\xi_2$ are essentially independent of system size $N$, suggesting
that finite-size effects on $\xi_2$ are rather weak.  Similar procedures 
are performed for the other localization lengths.

Our main results are presented in Fig.~2. All data have been obtained for
system size $N \geq 200$ and for the center of the band $E=0$. In view of 
the special nature of the doubly-occupied sites due to the on-site
interaction, it is natural to ask whether the definition for the
two-particle localization length $\xi_2$ correctly captures the CM
motion. To answer this question, we plot $\xi_{2,a}$ for interaction
strength $U=1.0$ as function of $a$ in Fig.\ 2(a). We find that
$\xi_{2,a}$ remains unchanged up to rather large $a$, implying that
$\xi_2$ is indeed a good description of the CM motion.  In fact,
$\xi_{2,a}$ remains independent of $a$ even for $a>\xi_1$. Hence, as 
opposed to the previous beliefs\cite{she,opp} the
interaction affects the two-particle motion even if the particle
distance exceeds $\xi_1$.  This can be understood in terms of
single-particle propagation in the 2d lattice of Fig.~1.  $\xi_{2,a}$
is associated with the transition probability along the dashed line a
distance $\sim a$ from the diagonal.  We recall that when $\langle
m,m-a|G|n,n-a\rangle$ is expanded in powers of the hopping matrix
element $t$, it is given by a sum over all possible paths from
$|n,n-a\rangle$ to $|m,m-a\rangle$.  If the distance between these two
sites, $\sim |m-n|$, is much smaller than $a$, the effect of the 
interaction would be
negligible. However, $\xi_{2,a}$ is defined by the limiting behavior
of $|m-n| \to \infty$ with $a$ finite, cf., Eq.~(5). In this case, the
contributions of paths which are sensitive to the interaction $U$ are
no longer negligible and $\xi_{2,a}$ remains influenced by the
interaction even though $a > \xi_1$.

In Fig.~2(b), we show our results for $\xi_r$ (symbols) which describes
the decay of the Green function with relative distance.  For comparison,
we also plot $\xi_2$ (lines).  At $U = 0$,
the two lengths are equal within the numerical accuracy, i.e., $\xi_2 
\simeq \xi_r$.  As $U$ increases, $\xi_r$ remains nearly constant 
while $\xi_2$ shows a pronounced enhancement in qualitative agreement 
with Shepelyansky's prediction\cite{she}.

In Fig.~2(c) we plot $\xi_f$ (symbols) and $\xi_2$ (lines), where the 
former describes the range over which one particle moves with the other 
one fixed.  At $U=0$, we find that $\xi_f$ approximately equals to
$2\xi_2$. As already seen in Fig.\ 2(b), $\xi_2$ shows a strong
increase with $U$. By comparison, $\xi_f$ shows a much weaker
increase. Hence, there exists a $U_c(W)$ beyond which $\xi_2$
exceeds $\xi_f$.

At $U=0$, the two particles move independently and the propagation of a
given particle is not affected by whether the other is moving in the
reverse ($\xi_r$) or in the same direction ($\xi_2$), implying
$\xi_r=\xi_2$.  Moreover, since $\xi_2$ measures two-particle
propagation, while $\xi_f$ the single-particle motion, one expects
that the transition probability for $\xi_2$ is given by the square of
that for $\xi_f$, so that $\xi_2 = \xi_f/2$. We note in passing that
$\xi_2(U=0)\neq\xi_1/2$, as was pointed out in ref.~\cite{song}.
Since both $\xi_r$ and $\xi_f$ are determined by the limiting behavior 
for diverging distance between the particles, one does not expect
them to be strongly influenced by the interaction. This explains the
rather flat dependences of both lengths on $U$.

With the additional information from $\xi_r$ and $\xi_f$, we can now 
construct a wavefunction picture in the 2d lattice representation of 
the problem (Fig.\ 1).  At $U=0$, this implies that the wavefunction 
profile is described by a square as shown by a thick solid line in 
Fig.~1(a).  As $U$
increases, the length of the edge associated with $\xi_2$
increases while that associated with $\xi_r$ remains
essentially constant. For $U>U_c$, the wavefunction profile becomes
highly anisotropic and we find that it can be well described by an
ellipse as shown by the thick solid line in Fig.~1(b).  The
elliptical shape predicts the relation
\begin{equation}
\xi_f = \frac{2\xi_2\xi_r}{\sqrt{\xi_2^2+\xi_r^2}}.
\end{equation} 
We have checked that our data are in good agreement with this expression for
$U>U_c$.  This clearly shows that the enhancement effect is associated
predominantly with the direction of $\xi_2$, i.e., the CM motion of
the two particles.

These results allow us to resolve some of the above mentioned
discrepancies between previous numerical studies.  We start by 
noting that the TMM measures
the largest length scale from the $N^2$ Green-function entries
$\langle 1n|G|Nm\rangle$ with $1 \leq m,n \leq N$\cite{fra2}.
According to our results, there are two competing lengths $\xi_2$ and
$\xi_f$.  For $U<U_c$, we find that the largest length is $\xi_f$ while 
for $U>U_c$, it is $\xi_2$.  Therefore, the TMM actually measures
$\xi_f$ for $U<U_c$ and $\xi_2$ only for $U>U_c$. We first compare our
results to those of Frahm {\it et al.}\cite{fra1}. We find that their
results for the localization length are two to three times larger than
our result for ${\rm max}\{\xi_f,\xi_2\}$ at given values of $W$ and
$U$. We attribute this to large finite-size effects in the TMM, as
suggested in ref.~\cite{rom}.  On the other hand, in ref.~\cite{rom}, any
enhancement effect was attributed to finite-size effects and it was
suggested that the TMM produces the single-particle localization
length for a sufficiently large system size. This is clearly
inconsistent with the results of the present paper. We
suggest the following explanation for the numerical results of ref.\
\onlinecite{rom}. The argument of ref.\ \onlinecite{rom} is based on
TMM data for $(W,U) = (3.0,1.0)$. For these parameters, we find that
the largest length is $\xi_f = 13.2 \pm 0.3$, which is close to $\xi_1
\simeq 11.7$. Hence, we expect that the data in ref.\
\onlinecite{rom} in fact extrapolate to $\xi_f$ which is
indistinguishable from $\xi_1$ within the numerical accuracy of
ref.~\cite{rom}. Therefore, we contend that the principal argument of
ref.~\cite{rom} is a misinterpretation of data for a special parameter
set and expect that the TMM exhibits the enhancement effect clearly
once $\xi_2 \gg \xi_f$. Finally, we find that the $\xi_2$'s in
ref.~\cite{opp} are somewhat larger than those in this paper, which we
attribute to the approximate treatment of the Green function in
ref.~\cite{opp}.

We thank H.A.\ Weidenm\"uller for useful discussions.


\noindent
\vspace{-0.7cm}
\begin{figure}
\centerline{\epsfxsize=7cm \epsfbox{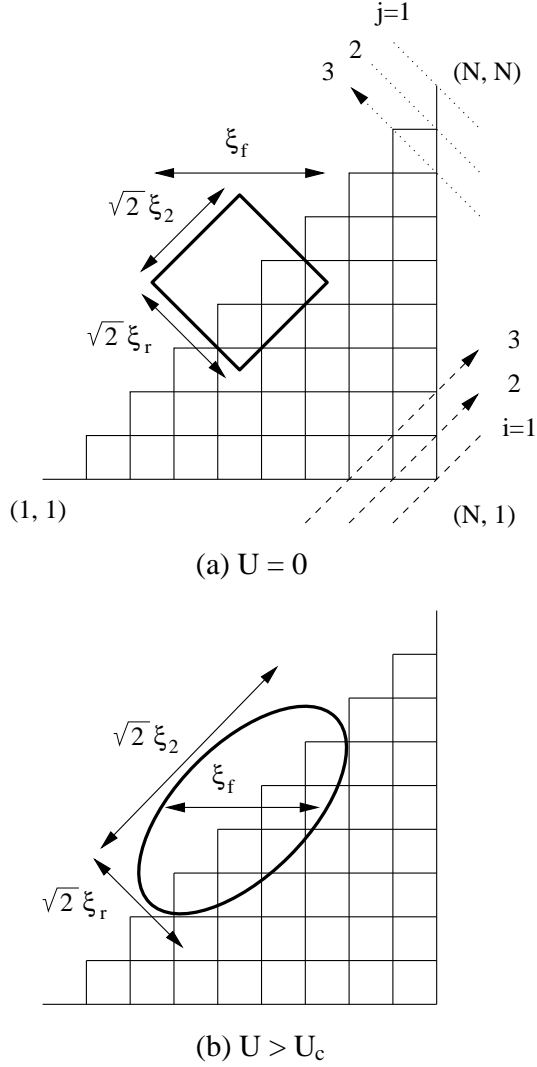}}
\vspace{4mm}
\caption{
Sketch of the two dimensional lattice (thin solid lines) and the wavefunction 
profile (thick solid lines).  The dashed lines (dotted lines)
represent the index scheme for the calculation of 
$\protect \xi_{2,a}$ $(\protect\xi_r)$.
Lengths are measured in terms of the lattice constant $d$.    
The factor $\protect \sqrt{2}$ arises because  $\protect\xi_r$ 
and $\protect\xi_2$ are defined in units of the diagonal length 
of the smallest square of the lattice ($\protect \sqrt{2} d$) while
$\protect\xi_f$ is defined in units of its edge length ($\protect d$).  
}
\end{figure}
\noindent
\vspace{-0.7cm}
\begin{figure}
\centerline{\epsfxsize=7.5cm \epsfbox{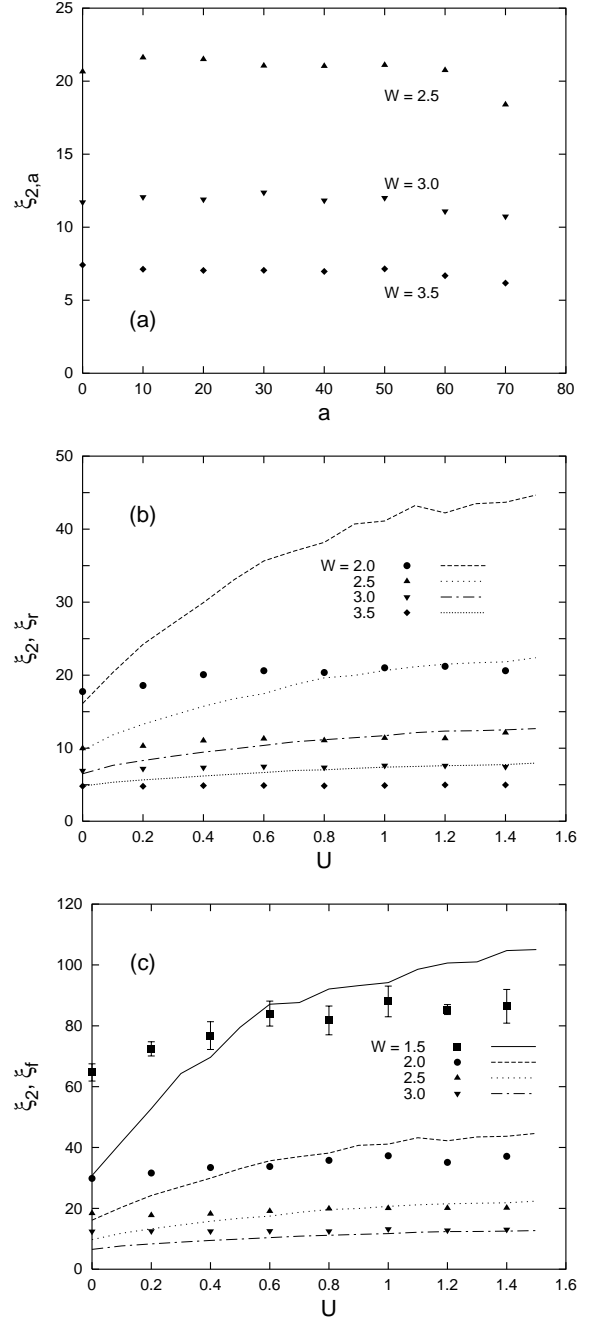}}
\vspace{4mm}
\caption{
(a) $\protect\xi_{2,a}$ as function of $\protect a$.  (b) 
$\protect \xi_2$ (lines) and
$\protect\xi_r$ (symbols) as function of 
$\protect U$.  (c) $\protect \xi_2$ (lines)
and $\protect\xi_f$ (symbols) as function of 
$\protect U$.  All data have been
obtained for system size $200 \protect\leq N \protect\leq 300$ and for 
$\protect E=0$.
}
\end{figure}

\end{multicols}
\end{document}